\documentclass[aps, prl, twocolumn, amssymb, amsmath, showpacs, superscriptaddress]{revtex4}

\usepackage{graphicx}
\usepackage{dcolumn}
\usepackage{bm}
\usepackage{xcolor}
\usepackage{ulem}
\usepackage{times}
\usepackage[colorlinks,citecolor=blue,linkcolor=blue]{hyperref}

\begin{document}

\title{Tailoring Magnetic Doping in the Topological Insulator Bi$_{2}$Se$_{3}$}

\author{Jian-Min Zhang}
\affiliation{School of Physics, Beijing Institute of Technology, Beijing 100081, China}
\affiliation{Beijing National Laboratory for Condensed Matter Physics and Institute
of Physics,\linebreak{} Chinese Academy of Sciences, Beijing 100190, China}

\author{Wenguang Zhu}
\email{wzhu3@utk.edu}
\affiliation{Department of Physics and Astronomy, The University of Tennessee,
Knoxville, Tennessee 37996, USA}
\affiliation{Materials Science \& Technology Division, Oak Ridge National Laboratory,
Oak Ridge, Tennessee 37831, USA}
\affiliation{ICQD/HFNL, University of Science and Technology of China, Hefei, Anhui, 230026, China}

\author{Ying Zhang}
\affiliation{Department of Physics, Beijing Normal University, Beijing 100875, China}

\author{Di Xiao}
\affiliation{Materials Science \& Technology Division, Oak Ridge National Laboratory,
Oak Ridge, Tennessee 37831, USA}
\affiliation{Department of Physics, Carnegie Mellon University, Pittsburgh, Pennsylvania 15213, USA}

\author{Yugui Yao}
\email{ygyao@bit.edu.cn}
\affiliation{School of Physics, Beijing Institute of Technology, Beijing 100081, China}
\affiliation{Beijing National Laboratory for Condensed Matter Physics and Institute
of Physics,\linebreak{} Chinese Academy of Sciences, Beijing 100190, China}

\begin{abstract}
We theoretically investigate the possibility of establishing ferromagnetism in the topological insulator Bi$_{2}$Se$_{3}$ via magnetic doping of $3d$ transition metal elements. The formation energies, charge states, band structures, and magnetic properties of doped Bi$_{2}$Se$_{3}$ are studied using first-principles calculations within density functional theory. Our results show that Bi substitutional sites are energetically more favorable than interstitial sites for single impurities. Detailed electronic structure analysis reveals that Cr and Fe doped materials are still insulating in the bulk but the intrinsic band gap of Bi$_{2}$Se$_{3}$ is substantially reduced due to the strong hybridization between the $d$ states of the dopants and the $p$ states of the neighboring Se atoms. The calculated magnetic coupling suggests that Cr doped Bi$_{2}$Se$_{3}$ is possible to be both ferromagnetic and insulating, while Fe doped Bi$_{2}$Se$_{3}$ tends to be weakly antiferromagnetic.

\end{abstract}
\pacs{71.20.Nr 61.72.U- 75.50.Pp}
\maketitle

The recent discovery of topological insulators (TIs) in narrow band gap semiconductors has generated great interest in both condensed matter physics and materials science~\cite{Bernevig2006,Moore2007,Fu2007a,Zhang2009,Xia2009,Hsieh2009,Hasan2010}. The strong spin-orbit coupling (SOC) in these materials, which is responsible for the nontrivial band topology, also highlights their application potential in semiconductor spintronics~\cite{Qi2011_RevModPhys}.  In particular, the intricate interplay between topological order and ferromagnetism is expected to give rise to a variety of unconventional spintronic effects that may lead to entirely new device paradigms~\cite{Liu2009,Qi2009,Garate2010,Wray2011b}. For example, a weak magnetic perturbation can open up an energy gap in the surface spectrum of a TI, resulting in the emergence of massive Dirac fermions~\cite{Chen2010i}.  In a TI thin film, long range magnetic order can also be established, which potentially could realize the long-sought quantum anomalous Hall state~\cite{Yu2010b,Xue-arXiv2011,Qiao2010}.

A natural strategy to introduce magnetism in TIs is via magnetic doping, similar to diluted magnetic semiconductors~\cite{Jungwirth2006}.  The current research mainly focuses on the Bi$_{2}$Se$_{3}$ family~\cite{Chen2010i,Yu2010b,Cha2010,Jin2011,Xue-arXiv2011,Liu2012}. However, existing experimental evidences have shown that the solubilities of magnetic impurities are normally very limited in Bi$_{2}$Se$_{3}$ family compounds~\cite{Dyck2002andChien2007,Cha2010,Chen2010i,Choi2005}, magnetic impurities tend to form clusters instead of being individually distributed in the materials, and dopants have uncertain valency~\cite{a-Larson2008}. On the magnetic ordering, ferromagnetism was reported in Mn and Fe doped Bi$_{2}$Te$_{3}$~\cite{Choi2005,Kulbachinskii2002,Choi2004,Hor2010d,Niu2011} and V, Cr and Mn doped Sb$_{2}$Te$_{3}$~\cite{Dyck2002andChien2007,Choi2005,Choi2004}, while a spin glass state was found in Mn doped Bi$_{2}$Se$_{3}$~\cite{Choi2005}. For Cr doped Bi$_{2}$Se$_{3}$, both antiferromagnetism~\cite{Choi2011} and ferromagnetism~\cite{Haazen2012} were observed, and the observations in Fe doped Bi$_{2}$Se$_{3}$ were also rather controversial~\cite{Kulbachinskii2002,Sugama2001,Salman2012,Choi2011}. From a growth point of view, it is imperative to choose proper magnetic elements and be able to precisely control the magnetic impurity distribution inside the host materials~\cite{Chen2010i}. This relies on comprehensive knowledge of the kinetic and energetic behaviors of magnetic impurities in TIs and consequently the electronic and magnetic properties. Yet, such knowledge is still lacking in the literature.

In this Letter, we investigate the feasibility of tailing Bi$_{2}$Se$_{3}$ to a ferromagnetic insulator via doping $3d$ transition metal (TM) elements, including V, Cr, Mn, and Fe, using first-principles calculations within density functional theory. We find that Bi substitutional sites are energetically more favorable than interstitial sites for single impurities of these elements, and the optimal growth conditions for TM doping are presented. The calculated band structures reveal that only Cr and Fe doped materials are insulating in the bulk but the intrinsic band gap of Bi$_{2}$Se$_{3}$ is substantially reduced due to the strong hybridization between the $d$ states of the dopants and the $p$ states of the neighboring Se atoms. Further investigation on the magnetic coupling between the dopants clarifies some experimental discrepancies and suggests that Cr doped Bi$_{2}$Se$_{3}$ is likely to be ferromagnetic, while Fe doped material tends to be weakly antiferromagnetic.

Our first-principles density functional theory calculations are performed using the pseudopotential plane-wave method with projected augmented wave~\cite{Blochl1994} potentials and Perdew-Burke-Ernzerhof-type generalized gradient approximation (GGA)~\cite{Perdew1996} for exchange-correlation functional, as implemented in the Vienna \textit{ab initio} simulation package (VASP)~\cite{Kresse1993a,Kresse1996}. A plane-wave energy cutoff of 300 eV is consistently used in all the calculations.  Unless mentioned otherwise, SOC~\cite{SOCnote} is taken into account in all calculations. Crystalline Bi$_{2}$Se$_{3}$ has a rhombohedral structure and its unit cell is composed of three weakly coupled quintuple layers (QLs). To investigate the behaviors of isolated magnetic dopants in bulk Bi$_{2}$Se$_{3}$, we employ a $2\times2\times1$ supercell containing 24 Bi and 36 Se atoms, as shown in Fig. 1 (a). A gamma-centered $7\times7\times2$ mesh of special \textit{k} points is adopted for integrations over the Brillouin zone. In calculating the formation energies~\cite{Zhang1991,VandeWalle2004} of charged defects, a uniform background charge is introduced to keep the supercell in charge neutral. With the lattice constants fixed at the experimental values of a=4.138 {\AA} and c=28.64 {\AA}, the internal coordinates of all the atoms are fully relaxed until the residual forces on each atom are less than 0.02 eV/{\AA}. All these parameters ensure the numerical errors in the calculated defect formation energies are less than 20 meV.

\begin{figure}
\begin{centering}
\includegraphics[width=3.4in]{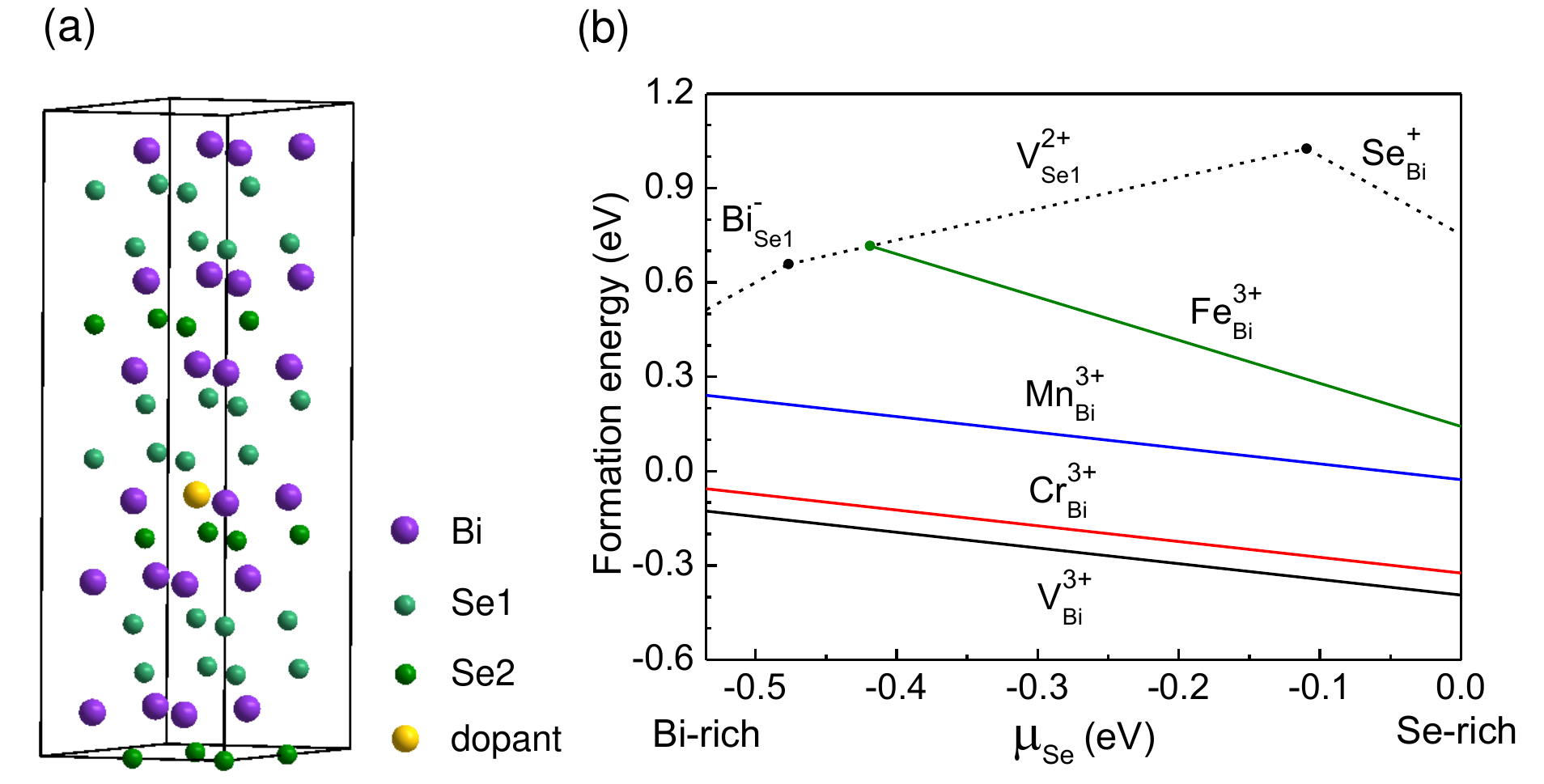}
\par\end{centering}

\centering{}\caption{(color online). (a) Illustration of a $2\times2\times1$  supercell for modeling a single dopant in bulk Bi$_{2}$Se$_{3}$. (b) Calculated formation energies of the most stable configurations of single V, Cr, Mn, and Fe impurities doped Bi$_{2}$Se$_{3}$ as a function of the host element chemical potentials. For comparison, the values of dominant intrinsic defects, selenium vacancies $V_{Se1}$, antisite defects $Bi_{Se1}$ and $Se_{Bi}$, are shown in dashed lines.}
\end{figure}

Extensive research on the traditional diluted magnetic semiconductors has indicated that their magnetic properties depend sensitively on the positions of magnetic impurities, namely, substitutional versus interstitial sites, in the host semiconductors~\cite{Jungwirth2006}. Therefore, we first examine the site preference of a serial of single $3d$ TM impurities (V, Cr, Mn, and Fe) in bulk Bi$_{2}$Se$_{3}$. To this end, we place the TM impurities at all possible interstitial and substitutional sites in bulk Bi$_{2}$Se$_{3}$ and compare their formation energies, which can be computed using the following expression~\cite{VandeWalle2004}
\begin{align}
\Delta H_{f}(TM) & =E_{tot}(TM)-E_{tot}(bulk)-\sum_{i}n_{i}\mu_{i},\label{eq:formation energy formula}
\end{align}
where $E_{tot}(TM)$ is the total energies of a supercell containing one impurity, $E_{tot}(bulk)$ is the total energy of the equivalent supercell containing only host atoms. $\mu_{i}$ denotes the chemical potential for species $i$ (host atoms or dopants), and $n_{i}$ indicates the corresponding number that have been added to or removed from the supercell. Given the host material being a binary compound, the values of $\mu_{i}$ are subject to the following relations: (i) $2\mu_{Bi}+3\mu_{Se}=\Delta H_{f}(Bi_{2}Se_{3})$, where $\Delta H_{f}(Bi_{2}Se_{3})$ is the formation energy of bulk Bi$_{2}$Se$_{3}$, to maintain equilibrium growth of Bi$_{2}$Se$_{3}$; (ii) $\mu_{Bi} \leqslant \mu_{Bi}^{0}$, $\mu_{Se} \leqslant \mu_{Se}^{0}$, and $\mu_{TM} \leqslant \mu_{TM}^{0}$ to avoid precipitation of Bi, Se, and TM elementary substances, where $\mu_{Bi}^{0}$, $\mu_{Se}^{0}$, and $\mu_{TM}^{0}$ refer to the chemical potentials of the stable Bi, Se, and TMs crystals, respectively. We chose rhombohedral Bi, hexagonal Se, nonmagnetic body-centered-cubic (bcc)  V, antiferromagnetic bcc Cr, antiferromagnetic face-centered-cubic Mn, and ferromagnetic bcc Fe as reference to evaluate the chemical potentials of the elements; (iii) $x\mu_{TM}+y\mu_{Se}\leqslant\Delta H_{f}(\text{TM}_{x}\text{Se}_{y})$ to ensure that all possible competing phases $\text{TM}_{x}\text{Se}_{y}$ can not precipitate.

\begin{figure}
\begin{centering}
\includegraphics[scale=0.6]{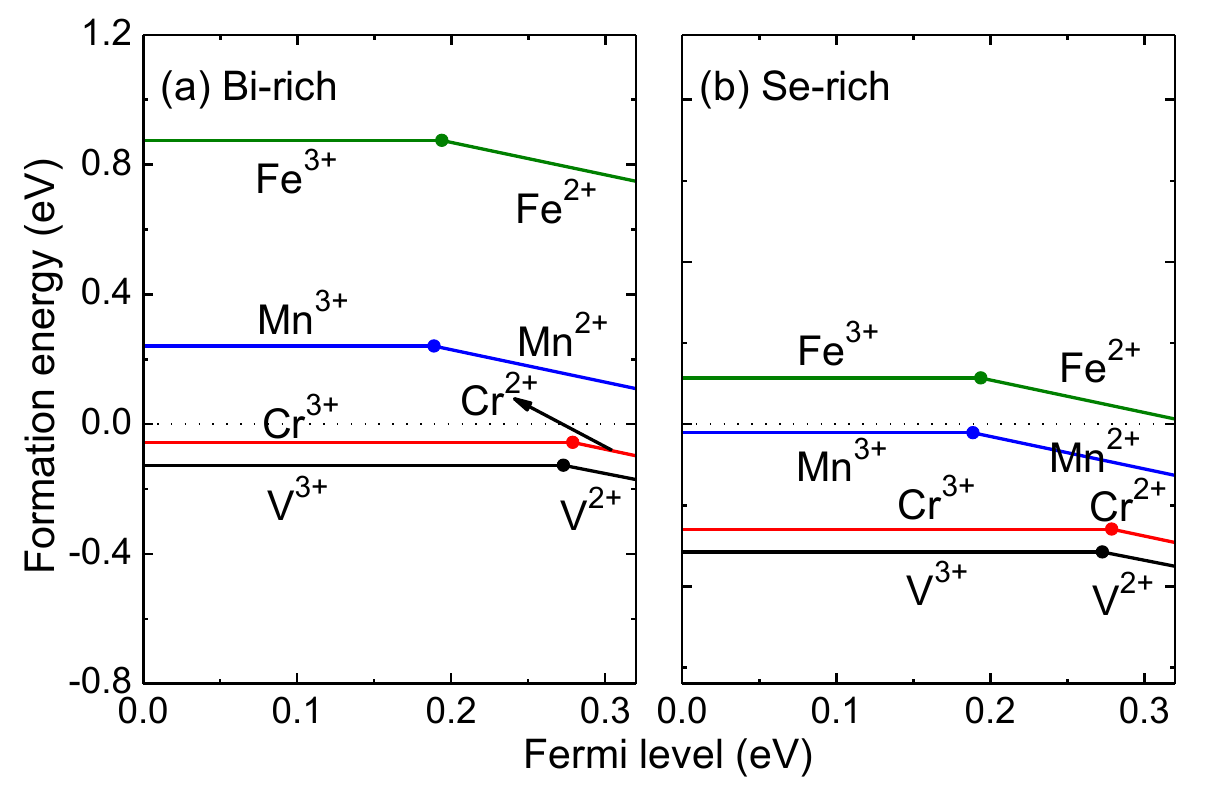}
\par\end{centering}

\centering{}\caption{(color online). Calculated formation energies of the most favorable charge states of single V, Cr, Mn, and Fe impurities doped at Bi substitutional sites in Bi$_{2}$Se$_{3}$ as a function of the Fermi level under (a) Bi-rich and (b) Se-rich conditions. }
\end{figure}

\begin{figure*}
\begin{centering}
\includegraphics[scale=0.80]{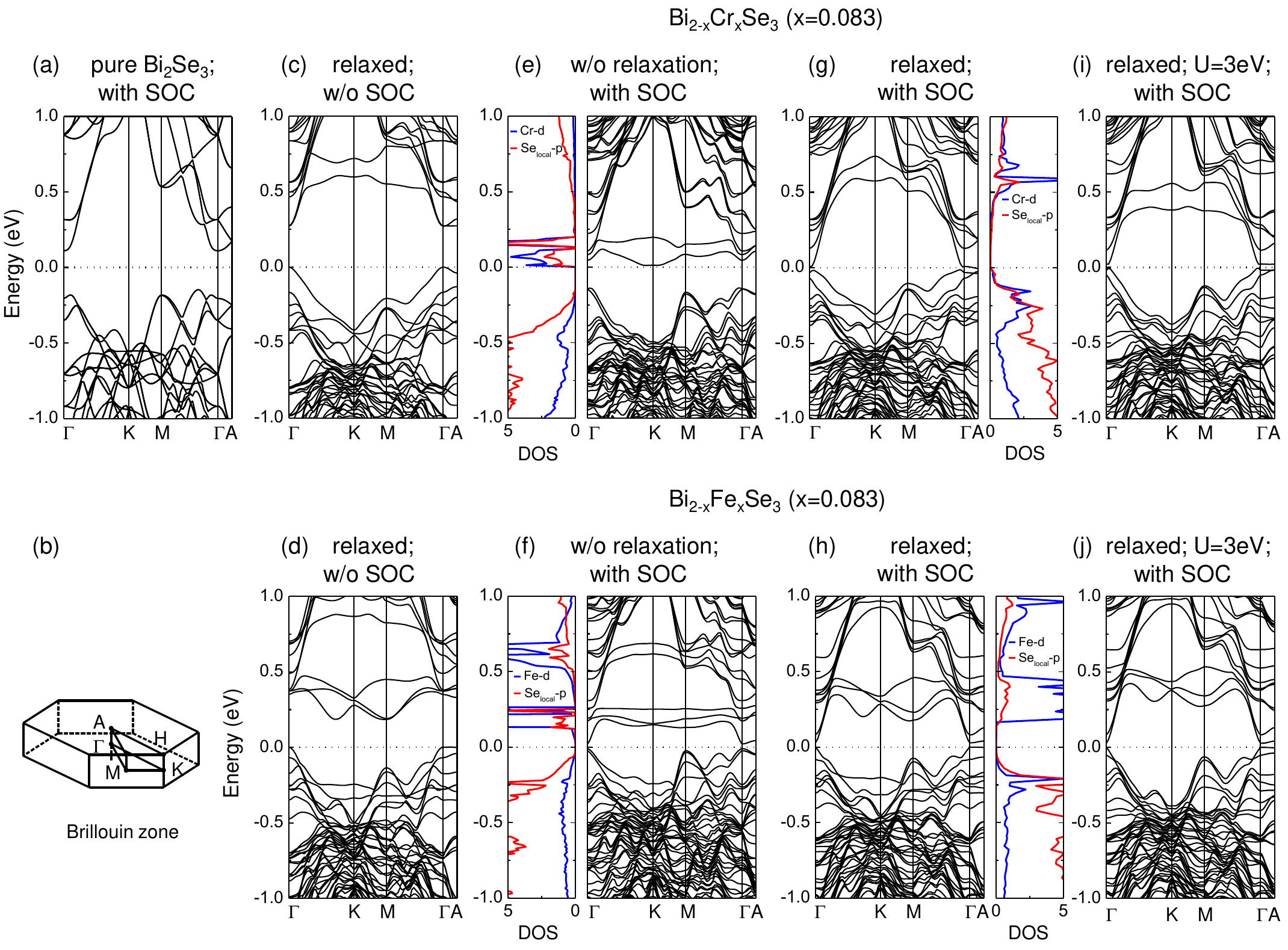}
\par\end{centering}

\centering{}\caption{(color online). (a) Band structure of Bi$_{2}$Se$_{3}$ calculated for a $2\times2\times1$ relaxed supercell with SOC. (b) Brillouin zone and high symmetry points of the $2\times2\times1$ supercell. (c)-(j) Band structures of Cr doped and Fe doped Bi$_{2}$Se$_{3}$ calculated with: (c) and (d) structural relaxation but without SOC; (e) and (f) SOC but without structural relaxation; (g) and (h) SOC and structural relaxation; (i) and (j) SOC, structural relaxation, and GGA+U (U=3 eV). Structural relaxations were all performed without the inclusion of SOC. Projected partial density of states (DOS) on the dopants and the neighboring Se atoms are also shown in (e)-(h).
}
\end{figure*}

Our calculated formation energies indicate that Bi substitutional sites are strongly preferred by all the TM impurities we investigated, compared to Se substitutional sites and all possible interstitial sites in bulk Bi$_{2}$Se$_{3}$. Interstitial sites in the van der Waals gap are the most favorable interstitial sites, while their formation energies are always larger than the dopants at substitutional Bi by at least 0.29 and 1.09 eV at Bi-rich and Se-rich conditions, respectively. The reported experimental observations of interstitial dopants in the van der Waals gap may be due to certain kinetic constraint, that is, the growth temperatures are not high enough for the dopants to overcome the energy barriers to incorporate into the substitutional sites. Fig. 1(b) shows the formation energies of the TM impurities at Bi substitutional sites as a function of the chemical potentials of Bi and Se. The order of the formation energies, $\Delta H_{f}(V) < \Delta H_{f}(Cr) < \Delta H_{f}(Mn) < \Delta H_{f}(Fe)$, can be attributed to the size effect of the elements, that is, the larger  the atomic radius of the impurity, the closer to that of the replaced Bi atom and thus the lower the formation energy. Among the four TM elements, V and Cr possess negative formation energies in the whole range of the accessible host element chemical potentials, suggesting that these two elements can be spontaneously introduced in bulk Bi$_{2}$Se$_{3}$.

To further identify the optimal growth conditions, we investigate the competition of the TM doping with the formation of intrinsic defects. The potential intrinsic defects include bismuth vacancies $V_{Bi}$, selenium vacancies $V_{Se}$, and antisite defects $Bi_{Se}$ and $Se_{Bi}$ in bulk Bi$_{2}$Se$_{3}$. The most favorable intrinsic defects and their formation energies are also shown in Fig. 1(b). We find that at extremely Bi-rich growth conditions, antisite defect $Bi_{Se1}$ is favorable, and Bi$_{2}$Se$_{3}$ is expected to be intrinsic $p$ type. However, at the other growth conditions, selenium vacancies $V_{Se1}$ and antisite defects $Se_{Bi}$, with both acting as donors, are dominant, rendering Bi$_{2}$Se$_{3}$ intrinsic $n$ type. This result provides a good explanation to experimental study~\cite{Xia2009,Hsieh2009} and guidelines for growth control. Our results also indicate that the formation energies of V, Cr, and Mn are always lower than those of the intrinsic defects, whereas selenium vacancies ($V_{Se1}$) and antisite defects ($Bi_{Se1}$) are energetically more favorable at Bi-rich conditions in the case of Fe doping. Therefore, in order to effectively dope Fe in bulk Bi$_{2}$Se$_{3}$, it is required to grow the materials at Se-rich conditions. However, the thermodynamic solubility of Fe in Bi$_{2}$Se$_{3}$ is expected to be limited, where the maximum doping concentration of Bi$_{2-x}$Fe$_{x}$Se$_{3}$ is estimated to be $x=2.24\%$ under Se-rich conditions at the melting point of Bi$_{2}$Se$_{3}$ of 979 K, according to Ref.~\cite{VandeWalle2004}. Experimentally, the effective doping concentration of Fe in Bi$_{2}$Se$_{3}$ is reported to be less than 2\%~\cite{Cha2010}, which is consistent with our result.

In the above calculations, the whole systems are kept charge neutral, and the TM impurities, therefore, hold a $3+$ charge state as they are introduced in Bi sites in Bi$_{2}$Se$_{3}$.  Next, we study the formation energies of the doped TM impurities in various charge states. In this case, the definition of the impurity formation energy is modified as \cite{VandeWalle2004}
\begin{align}
\Delta H_{f}(TM,\, q) & =E_{tot}(TM,\, q)-E_{tot}(bulk)-\sum_{i}n_{i}\mu_{i}\nonumber \\
 & +q(E_{F}+E_{V}+\triangle V),\label{eq:formation energy formula1}
\end{align}
where $E_{tot}(TM,\, q)$ is the total energy of a supercell containing one impurity in charge state $q$ and $E_{F}$ is the Fermi level with respect to the valence band maximum of the intrinsic bulk Bi$_{2}$Se$_{3}$ ($E_{V}$). $\triangle V$ is a potential alignment due to different energy reference among the defect supercell and pure supercell. Figs. 2(a) and 2(b) show the calculated formation energies of the TM impurities in their favorable charge states as a function of  $E_{F}$ at Bi-rich and Se-rich growth conditions, respectively. The $3+$ charge state of the TM impurities are preserved when $E_{F}$ is smaller than 0.27, 0.28, 0.19, and 0.19 eV for V, Cr, Mn, and Fe doped cases, respectively. At extreme $n$-type conditions, i.e., $E_{F}$ approaching the conduction band minimal, the charge state of the TM impurities changes to $2+$, which may result in the change of carrier concentration in the materials when doping TM atoms. From Figs. 2(a) and 2(b), we obtain that the Se-rich condition is more optimal than the Bi-rich condition for doping TM in Bi$_{2}$Se$_{3}$. A systematic investigation of the molecular beam epitaxy growth for Bi$_{2}$Se$_{3}$ films also indicates that the optimal growth can be achieved under a Se-rich atmosphere~\cite{Song2010}, which agrees with our results.

We next focus our attention on the electronic structures of the TM doped Bi$_{2}$Se$_{3}$. Structural relaxations were all performed without the inclusion of SOC. The calculated band structures do not show any noticeable difference, in comparison with the calculations with SOC included in the relaxation for Cr and Fe doped Bi$_{2}$Se$_{3}$, and the band gap differences are less than 0.01 eV. Figure 3(a) shows the calculated band structure of pure Bi$_{2}$Se$_{3}$ with SOC, which gives a band gap of 0.32 eV. In the TM doped cases, even without SOC, Cr, and Fe doping results in insulating magnetic states with energy band gaps of 0.28 [Fig. 3(c)] and 0.18 eV [Fig. 3(d)], respectively, whereas V and Mn create impurity states in the band gap of Bi$_{2}$Se$_{3}$, which completely close the band gap and make the materials metallic. With SOC, we obtain band gaps of 0.15 and 0.03 eV for Cr and Fe doped Bi$_{2}$Se$_{3}$ as the dopants are fixed at the Bi sites without any structural relaxation [Figs. 3(e) and 3(f)]. However, after the structural relaxation, the band gaps are further reduced, as shown in Figs. 3(g) and 3(h), to 0.010 and 0.028 eV for Cr and Fe doped Bi$_{2}$Se$_{3}$, respectively. This seemingly counterintuitive result can be attributed to structural relaxation induced enhancement of the electronic state hybridization between the TM dopants and their neighboring Se atoms. When all the atoms are fixed at their original crystal positions, due to the relatively small covalent radius of Cr and Fe compared with Bi, Cr and Fe states have little overlap with the neighboring Se atoms, resulting in atomiclike localized electronic states, as evidenced by the presence of the fairly flat impurity bands in the gap region shown in Figs. 3(e) and 3(f). After the structural relaxation, the distance between Cr (Fe) and the neighboring Se is decreased by about 0.4 {\AA} (0.3 {\AA}) and the impurity bands are substantially broadened, suggesting strong  hybridization between the TM dopants and the neighboring Se atoms. A more detailed analysis on the projected partial density of states reveals that the $d$ states of Cr and Fe hybridize the $p$ states of Se.

We further performed GGA+U calculations with U on the TM impurities ranging from 3 to 6 eV and J=0.87 eV to describe the strong electron-electron correlation in partially filled $3d$ TM elements. The results, as shown in Figs.~3(i) and 3(j), indicate that the band gaps of Cr and Fe doped Bi$_{2}$Se$_{3}$ increases to 0.025 and 0.028 eV for U=3 eV and 0.026 and 0.028 eV for U=6 eV, respectively. Thus, electron-electron correlation could slightly enlarge the band gaps.

Finally, we discuss the magnetic properties of the TM doped Bi$_{2}$Se$_{3}$. The magnetic ground states for single impurity doped Bi$_{2}$Se$_{3}$ have been calculated with the inclusion of SOC, the results indicate that both Cr and Fe prefer the direction perpendicular to the Bi$_{2}$Se$_{3}$ QLs, with the magnetic moments close to 3 and 5 $\mu_{B}$, respectively. Our results are different than the recent reports of in-plane magnetization easy axis in Ref.~\cite{Honolka2012}, where Fe adatoms on the surface of Bi$_{2}$Se$_{3}$(111) and the intergrown composite crystal of Fe$_{7}$Se$_{8}$ and Bi$_{2}$Se$_{3}$~\cite{Ji2012}, in which magnetic anisotropy is attributed to the intrinsic properties of ferromagnet Fe$_{7}$Se$_{8}$. The magnetic coupling between the TM dopants is estimated by placing two TM dopants in the supercell at various separation distances and calculating the total energy difference between antiferromagnetic (AFM) and ferromagnetic (FM) states of the dopants at the same distance. Our preliminary results reveal that the magnetic coupling dominantly favors FM in the Cr doped Bi$_{2}$Se$_{3}$ while exclusively favors weak AFM in the Fe doped case. In the Cr doped case, the coupling strength [$(E_{AFM}-E_{FM})/2$] is on the order of 10 meV for the two Cr atoms within the same QL at the first three nearest neighboring distances and substantially reduced to the order of 1 meV or even less for the two Cr atoms in different QLs. We notice that studies of the competition between weak Localization and weak antilocalization from Lu \textit{et al}~\cite{Lu2011} indicated that a ferrromagnetic order can be formed in Cr doped Bi$_{2}$Se$_{3}$, which was later verified experimentally with a nominal Cr doping level of $23\%$~\cite{Liu2012}. More recently, ferromagnetism was reported to be observed in Cr doped Bi$_{2}$Se$_{3}$~\cite{Haazen2012}, which agrees well with our result.

In the Fe doped case, AFM is favorable while its coupling strength is very weak (on the order of 1 meV), even though the distance of two Fe atoms is close to 9.84 {\AA}. Experimentally, as we previously discussed, the doping concentration of Fe in Bi$_{2}$Se$_{3}$ is found to be lower than $2\%$~\cite{Cha2010}, where the average distance between two uniformly distributed Fe atoms is about 11.79 {\AA}. The corresponding coupling strength is predicted to be even less than 1 meV in this case. Hence, AFM in Fe doped Bi$_{2}$Se$_{3}$ may not be obvious to detect, especially in the low doping concentration regime, which agrees with experimental observations of nonmagnetism~\cite{Kulbachinskii2002,Sugama2001}. In addition, heavily Fe doped Bi$_{2}$Se$_{3}$ often exhibit FM behavior~\cite{Choi2011}, since Fe often tends to form clusters or become phase separated instead of being uniformly distributed during growth.

In summary, we have studied the effects of magnetic doping of a series of $3d$ transition metal elements in the TI Bi$_{2}$Se$_{3}$ using first-principles calculations.  Our calculated formation energies indicate that Bi substitutional sites are strongly preferred by all the TM impurities.  By examining the energetics of the dopants in the host material and the resulting electronic structures and magnetic properties, we have found that Cr and Fe doping preserves the insulating nature of the host TI in the bulk but the intrinsic band gap is substantially reduced due to the strong hybridization between the electronic states of the dopants and the neighboring Se atoms. The further investigation on the magnetic coupling between the dopants has suggested that Cr doped Bi$_{2}$Se$_{3}$ is likely to be FM, while Fe doped material tends to be weak AFM. Only Cr doped Bi$_{2}$Se$_{3}$ with both ferromagnetic and insulating properties is promising for realizing the quantized anomalous Hall effect.

J.Z. and Y.Y. were supported by the MOST Project of China (Grant No. 2011CBA00100), NSF of China (Grants No. 10974231, No. 11174337, and No. 11225418). W.Z. and D.X. were supported by the U.S. Department of Energy, Office of Basic Energy Sciences, Materials Sciences and Engineering Division. W.Z. was also supported by the National Natural Science Foundation of China (grant No. 11034006). The calculations were performed at the Supercomputing Center of Chinese Academy of Sciences.


\begin{thebibliography}{100}
\bibitem{SOCnote} Due to the important relativistic effects on Bi$_{2}$Se$_{3}$, SOC was taken into account in our calculations. The test value of formation energy for Bi$_{2}$Se$_{3}$ with SOC is about 17\% different from that without SOC.
\end{thebibliography}
\end{document}